\RequirePackage{fix-cm}
\documentclass[smallextended,referee,nospthms]{svjour3}       
\smartqed  

\title{A toolbox of Equation-Free functions in Matlab\slash Octave for efficient system level simulation}
\titlerunning{A toolbox of Equation-Free functions}

\author{John~Maclean
\and J.~E.~Bunder
\and A.~J.~Roberts
}

\institute{%
John~Maclean \at School of Mathematical Sciences, University of Adelaide, South Australia.
\url{http://www.adelaide.edu.au/directory/john.maclean}%
\and{%
J.~E.~Bunder \at School of Mathematical Sciences, University of Adelaide, South Australia.
\url{mailto:judith.bunder@adelaide.edu.au},
\url{http://orcid.org/0000-0001-5355-2288}}%
\and{%
A.~J.~Roberts \at School of Mathematical Sciences, University of Adelaide, South Australia.
\url{http://www.maths.adelaide.edu.au/anthony.roberts},
\url{http://orcid.org/0000-0001-8930-1552}}}

\date{\today}

\usepackage{natbib}
\usepackage[T1]{fontenc}
\usepackage[dvipsnames]{xcolor}
\definecolor{matlab1}{rgb}{0,0.4470,0.7410}
\definecolor{matlab2}{rgb}{0.8500,0.3250,0.0980}
\definecolor{matlab3}{rgb}{0.9290,0.6940,0.1250}
\definecolor{matlab4}{rgb}{0.4940,0.1840,0.5560}
\definecolor{matlab5}{rgb}{0.4660,0.6740,0.1880}
\definecolor{matlab6}{rgb}{0.3010,0.7450,0.9330}
\definecolor{matlab7}{rgb}{0.6350,0.0780,0.1840}
\usepackage{algpseudocode,algorithm}
\algrenewcommand\textproc{\texttt}

\usepackage{pgfplots} 
\pgfplotsset{compat=newest} 
\usepgfplotslibrary{external} 
\usetikzlibrary{decorations.markings}
\usetikzlibrary{shapes,arrows,fit}
\usetikzlibrary{positioning,plotmarks}

\def\figurename{\em Figure}

\usepackage{fancyvrb}
\newenvironment{matlab}%
    {\Verbatim}
    {\endVerbatim}
\makeatletter
\def\fancyvrbStartStop{%
  \edef\FancyVerbStartString{\@percentchar\@charrb} 
  \edef\FancyVerbStopString{\@percentchar\@charlb} }
\makeatother

\usepackage{url,microtype,amsmath,amssymb,defns,graphicx,hyperref,doi}
\hypersetup{colorlinks,linkcolor=RoyalBlue,citecolor=RoyalBlue,pagecolor=RoyalBlue,%
            urlcolor=magenta,filecolor=magenta,breaklinks,%
            dvips,bookmarks,bookmarksopen}
\makeatletter
\AtBeginDocument{{\def\and{, }\def\thanks#1{}%
  \hypersetup{
    pdfauthor={\@author},
    pdftitle={\@title}}}
    }
\makeatother
\usepackage{mycleveref}

\def\into{\(\leftrightarrow\)\ }
\Vec f \Vec x \Vec u \Vec X \Vec U \Vec v \Vec b
\Bb R  \Bb X 

\makeatletter
\def\@oddfoot{\hfill\tiny\sf \today}
\makeatother

\graphicspath{{Figs/}}

\atBegin{algorithm}{\mathcode`\,="213B}

\begin{document}
\epstopdfsetup{suffix=} 

\maketitle

\begin{abstract}
The `equation-free toolbox' empowers the computer-assisted analysis of complex, multiscale systems.
Its aim is to enable you to immediately use microscopic simulators to perform macro-scale system level tasks and analysis, because micro-scale simulations are often the best available description of a system.
The methodology bypasses the derivation of macroscopic evolution equations by computing the micro-scale simulator only over short bursts in time on small patches in space, with bursts and patches well-separated in time and space respectively.
We introduce the suite of coded equation-free functions in an accessible way, link to more detailed descriptions, discuss their mathematical support, and introduce a novel and efficient algorithm for Projective Integration. 
Some facets of toolbox development of equation-free functions are then detailed.
Download the toolbox functions\footnote{\protect\url{https://github.com/uoa1184615/EquationFreeGit}} and use to empower efficient and accurate simulation in a wide range of your science and engineering problems.  
\end{abstract}

\begin{keywords}
\ Multiscale methods $\cdot$ code toolbox $\cdot$ numerical algorithms
\end{keywords}

\tableofcontents

\section{Introduction}

Suppose that you have a \emph{detailed and trustworthy} computational simulation of some problem of interest. 
When the detailed computation is too expensive to simulate all the times of interest over all the space of interest, then the `Equation-Free Methodologies' aim to accurately empower long-time simulation and system level analysis \cite[e.g.,][]{Kevrekidis09a, Kevrekidis04a, Kevrekidis03b}. 
Our toolbox provides you with these methodologies coded into \script\ functions.

\subsection{Simulation on only small patches of space}
\label{sec:flagPat}

\begin{figure}
\centering 
\caption{\label{fig:patEx} 
Snapshots in time of the patch scheme simulation of the nonlinear diffusive micro-scale system~\cref{eq:pat}. }
\begin{tabular}{@{}ll@{}}
(a) \(t=0\)&(b) \(t=0.33\)\\
    \tikzsetfigurename{Figs/cP2t0}%
    \input{Figs/cP2t0}
     & %
    \tikzsetfigurename{Figs/cP2t1_3}%
    \input{Figs/cP2t1_3}
     \\
(c) \(t=1\)& (d) \(t=3\)\\
    \tikzsetfigurename{Figs/cP2t1}%
    \input{Figs/cP2t1}
     & %
    \tikzsetfigurename{Figs/cP2t3}%
    \input{Figs/cP2t3}
    
\end{tabular}
\end{figure}

\cref{fig:patEx} illustrates an `equation-free' computation on only small well-separated patches of the spatial domain.
The micro-scale simulations within each patch, here the nonlinear diffusive system~\cref{eq:pat}, are craftily coupled to neighbouring patches and thus interact to provide accurate macro-scale predictions over the whole spatial domain \cite[e.g.,][]{Roberts2011a}. 
We have proved that the patches may be tiny, and still the scheme makes accurate macro-scale predictions \cite[]{Roberts06d}.  
Thus the computational savings may be enormous, especially when combined with projective integration (\cref{sec:flagPI}).

The example system illustrated in \cref{fig:patEx} is a nonlinear discrete diffusion system inspired by the lubrication flow of a thin layer of fluid, namely
\begin{align*}
\D tu &=\divv(3u^2\grad u),
\qquad\text{equivalently}\quad \D tu=\delsq(u^3), 
\end{align*}
which on a 2D micro-scale lattice~$x_{i,j}$ with tiny micro-scale spacing~$d$ is here discretised simply to
\begin{align}
\label{eq:pat}
\de{t}{u_{i,j}} &= \frac{u_{i+1,j}^3
+u_{i-1,j}^3 +
u_{i,j+1}^3 +u_{i,j-1}^3 -4u_{i,j}^3 }{d^2}\,.
\end{align}
We want to predict the dynamics of this spatial micro-scale lattice system on the macro-scale spatial domain $[-2,\,2]\times[-3,\,3]$, but suppose full direct computation is too expensive.
Instead, the micro-scale simulation illustrated by \cref{fig:patEx} was performed only on about~20\% of the domain (it could be much less)---the small patches of space in \cref{fig:patEx}.
The key to an accurate macro-scale prediction is that each patch is coupled to nearby patches, at every computed time, by appropriate macro-scale interpolation that gives the edge values for every patch \cite[e.g.,][]{Roberts2011a}.

The patch scheme is most useful in applications where there is no known macro-scale closure.
Then the patch scheme automatically achieves a computational macro-scale closure, without the need for any analytic construction often invoked in numerical\slash computational homogenization \cite[e.g.,][]{Saeb2016, Geers2017, Peterseim2019}---the patch scheme is `equation-free'.  
Our approach could be classed as a dynamic homogenization \cite[e.g.,][]{Craster2015}.
Frequently, problems of interest in applications compute on a micro-scale spatial lattice as in the spatial discretization~\eqref{eq:pat}.
Suppose \(\xv_i\) are coordinates of a micro-scale lattice, for potentially exhaustingly many lattice points indexed by~$i$; for example, a full atmospheric simulation. 
And suppose your detailed and trustworthy simulation is coded in terms of micro-field variable values~\(\uv_i(t)\in\RR^p\) at lattice point~\(\xv_i\) at time~\(t\).
When a detailed computational simulation is prohibitively expensive over all the desired spatial domain, \(\xv\in\XX\subset\RR^{d}\), our toolbox provides functions that empower you to use your micro-scale code as a `black-box' inside only small, well-separated, patches of space by appropriately coupling across un-simulated space between the patches (\cref{sec:justPat}). 
The toolbox functions have many options including both newly developed spectral coupling, and new symmetry preserving coupling.  
\cref{sec:tutPat} gives an introductory tutorial.


\subsection{Projective Integration skips through time}
\label{sec:flagPI}

\begin{figure}
\centering
\caption{\label{fig:PIGsing}Projective Integration by the new function
\PIG\ of the example multiscale system~\cref{eq:sing}.
The macro-scale solution~\(U(t)\) is represented by the blue circles ({{\color{matlab1}$\boldsymbol{\circ}$}}). The black dotted line, underneath the PI solution, shows an accurate micro-scale simulation over the whole time domain. 
}
    \tikzsetfigurename{Figs/PIGsing}%
    \input{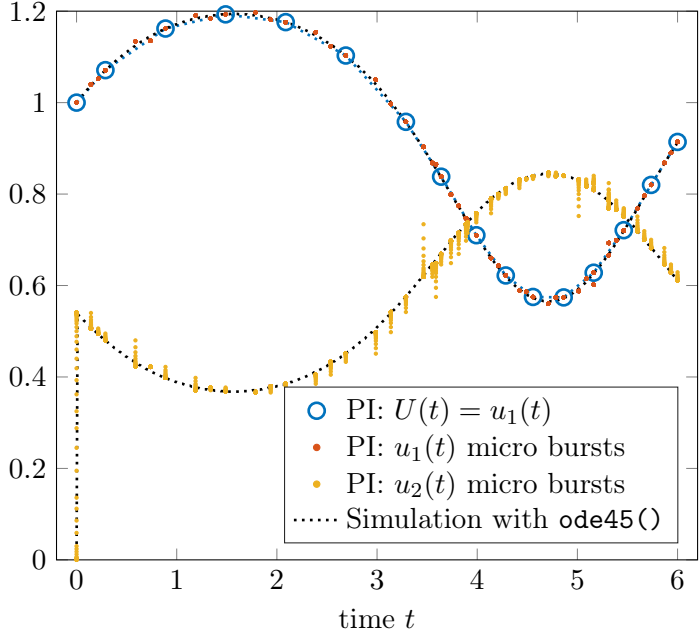}
    
\end{figure}

Simulation over time is a complementary dynamic problem.
The `equation-free' approach is to simulate for only short bursts of time, and then to extrapolate over un-simulated time into the future, or into the past, or perform system level analysis \cite[e.g.,][]{Gear02b, RicoMartinez2004, Erban2006, Givon06}.
\cref{fig:PIGsing} plots one example where the gaps in time show the un-computed times between bursts of computation. 

In general such a simulation is coded in terms of detailed (micro-scale) variable values~\(\uv(t)\), in~\(\RR^{p}\) for some~\(p\) (typically large in applications), and evolving in time~\(t\).
The details~\uv\ could represent particles, agents, or states of a system.
Both forward and backward in time computations may be performed by Projective Integration~(PI) with provable accuracy \cite[]{Gear02b, Givon06, Maclean2015}.
For efficient simulation on long times, \cref{sec:justPI} describes how to provide your micro-scale detailed \script\ code as a `black box' to the novel Projective Integration functions in the toolbox. 
\cref{sec:tutPI} gives a user-friendly introductory tutorial.

The example simulation of \cref{fig:PIGsing} is that of a toy system that nonetheless has challenging qualities of the multiscale phenomena that Projective Integration resolves. 
Here the micro-scale simulation is a pair of coupled slow-fast \ode{}s for \(\uv(t) = (u_1(t), u_2(t)) \): 
\begin{subequations} 
\label{eq:sing}
\begin{align}
\label{s1}\frac{du_1}{dt}=&\cos(u_1)\sin(u_2)\cos(t) ,\\
\label{s2}\frac{du_2}{dt}=&10^5\big[\cos(u_1)-u_2\big].
\end{align} 
\end{subequations}
Using simple integration schemes, numerical solutions can be rapidly computed on micro-times of \Ord{10^{-5}}, but solutions over \Ord{1} times are computationally prohibitive---except by stiff integrators. 
\cref{sec:choPI} discusses scenarios where stiff integrators cannot be used or are relatively expensive, but where projective integration is effective.

The system~\eqref{eq:sing} represents the realities of, for example, molecular dynamics simulations with rapid modes represented by~$u_2(t)$ and slow macro-scale state variables (like temperature) represented by~$u_1(t)$.
In applications the dynamics of the slow modes are usually not known, instead they emerge over the micro-scale simulation bursts \cite[e.g.,][]{Cisternas03, Setayeshgar05, Erban2006}.
Consequently, although in this toy system the slow variable~\(u_1\) and fast variable~\(u_2\) are obvious, here we compute only with the full system~\cref{eq:sing}---it is a `black-box' for which we do not necessarily know what `variables' are fast or slow.
An alternative is to invoke algebraic analysis to construct the slow manifold of slow-fast systems like~\eqref{eq:sing} \cite[e.g.,][Ch.~4--5]{Roberts2014a}: however, in many scenarios such analysis is not feasible.
Since one often only measures macro-scale state variables, here we suppose the micro-simulator only outputs $U(t) = u_1(t)$, called a restriction.
Projective Integration (PI) uses only short bursts of the `black-box' simulation, and then invokes an appropriate extrapolation to accurately predict large `projective' steps.
After a projective step the PI algorithm appropriately re-initialises~$u_2(t)$, called lifting, for another burst of the micro-simulator. 

\cref{fig:PIGsing} shows the PI computation and the associated micro-scale bursts, compared to expensive simulation with \texttt{ode45()}.
The micro-scale simulations with~\(u_2(t)\) show the characteristic fast transients.
The absolute error between the Projective Integration simulation and the trusted \texttt{ode45()} simulation is~\(2\cdot10^{-6}\).

The PI simulation of \cref{fig:PIGsing} was performed by a novel function, called \verb|PIG|, introduced in \cref{sec:justPI} and included in the toolbox.
The advantage of the Projective Integration simulation over \Matlab's \verb|ode45()| is that PI uses only~\(0.6\%\) of the number of evaluations of the \ode{}s~\cref{eq:sing}.
\cref{sec:tutPI} discusses how this new PI scheme and function may be constructed through recursive use of \verb|ode45()| which inherits all of \Matlab's adaptive error control---an error control active on both the micro-scale and \text{the macro-scale.}

\section{Equation-free algorithms}
\label{sec:just}
This section outlines the key Projective Integration and Patch Dynamics algorithms implemented in the toolbox \cite[]{Roberts2019b}. 
Pseudo-code highlights the essential features of each algorithm and the accompanying discussion demonstrates the extended capabilities. 
Theoretical support for the algorithms is also discussed.

\subsection{Algorithms for projective integration in time}
\label{sec:justPI}
Our efficient simulation of the stiff dissipative system \cref{eq:sing} is due to the class of algorithms called Projective Integrators.
Small time steps with the micro-simulator are alternated with long time steps comprised of `projective' extrapolations.
Sometimes PI is done with the macrostep being a Forward Euler method \cite[e.g.]{Siettos2003,Chuang15}, and so incurs global error proportional to the size of the projective time steps.
Such low accuracy motivated the development of PI algorithms of provably higher accuracy \cite[]{RicoMartinez2004,LeeGear07}.
Here we additionally extend the methodology to a projective integration that invokes adaptive integrators.

The toolbox provides \verb|PIG()|---denoting a Projective Integration with General macro-integrator and micro-simulator.
The user specified macro-integrator is any integrator suitable for time stepping on the macro-scale, and includes adaptive integrators such as \Matlab's \verb|ode45| as used for \cref{fig:PIGsing}.
Then \verb|PIG()| provides to that integrator accurate time derivatives at the required times---time derivatives estimated from appropriate relatively short bursts of the user-defined micro-simulator~\verb|microSim()|.
The effect is to do PI with any integrator, explicit or implicit, taking the projective steps.
Error analyses suggested that schemes of this sort usually incur significant errors proportional to the duration of the micro-simulator burst \cite[]{E03, Maclean2015, Maclean2015b}.
However, such errors are avoided by \verb|PIG()| through a novel implementation of a `constraint-defined manifold computing' scheme based on a methodology originally proposed by \cite{Gear04, Gear05}. 
\cref{alg:PIG} provides pseudo-code for \verb|PIG()| that details these steps. 
This package is the first time such functionality for Projective Integration has been developed into a general function, tested, \text{and made available.}

\begin{algorithm}[btp]
\caption{\label{alg:PIG}The \PIG() function uses short bursts of a user provided micro-scale process to simulate from time~$T_0$ to~$T$, with initial condition~$\uv_0$, and via a user-specified macro-scale integrator \mac().}

PIG tells the specified macro-scale integrator to use time derivatives estimated by a constraint-defined manifold function.
\begin{algorithmic}[1]
\Function{\PIG}{$\mac(), \mic(), \protect\UseVerb{proj}(), \{T_0, T\}, \uv_0$}
\State \Return $\{T_n, \uv_n \} _{n=0}^N := \mac(\protect\UseVerb{proj}(), \{T_0, T\}, \uv_0);$
\EndFunction
\end{algorithmic}

Here, the toolbox's constraint-defined manifold function uses two applications of $\Call{\mic}{}$ and a backwards projective step in order to provide the  time derivative at the time specified by \mac().

\begin{algorithmic}[1]
\Function{constrDeriv}{$t_0, \uv_0, \delta$}
\State $\{t_m,\uv_m\}_{m=0}^{M} := \Call{microSim}{t_0, \uv_0 , \delta};$ \quad(burst of simulation)
\State $ \left(\frac{\Delta \uv}{\Delta t}\right)_M := \frac{\uv_M - \uv_{M-1}}{t_M - t_{M-1}} ; \quad\text{(derivative estimate at } t_0+\delta )$
\State $\vv_0  := \uv_m - 2\delta\left(\frac{\Delta \uv}{\Delta t}\right)_M ; \quad\text{(PI backwards to } t_0-\delta )$
\State $\{s_m,\vv_m\}_{m=0}^{M} := \Call{microSim}{t_0-\delta, \vv_0, \delta} ;$ \quad(burst of simulation)
\State \Return $\left(\frac{\Delta \vv}{\Delta s}\right)_M := \frac{\vv_M - \vv_{M-1}}{ s_M -  s_{M-1}}; \quad\text{(derivative estimate at } t_0)$
\EndFunction
\end{algorithmic}

A user provides a micro-scale simulator with the following input and output.
\begin{algorithmic}[1]
\Function{microSim}{$t_0,  \uv_0, \delta$}
\State compute a burst with end time $t_M := t_0+\delta$;
\State \Return  $\{t_m,\uv_m\}_{m=0}^{M}$ 
\EndFunction
\end{algorithmic}

\end{algorithm}

\cref{alg:PIG} outlines the essence of \verb|PIG()|, but additional features may be invoked by a user.
In particular, sometimes the projective steps are performed on a few macro-scale variables only.
That is, the PI is done in a space of reduced dimension to that of the micro-simulator  \cite[e.g.]{Frederix2007, Bold2012, Sieber2018}.
In such cases the user provides `lifting' and `restriction' operators to convert between the micro- and macro-simulation spaces \cite[]{Roose2009}.

In addition to \verb|PIG()|, the toolbox provides efficient integrators \verb|PIRK2()| and \verb|PIRK4()|, which are Projective Integrators similar to \verb|PIG()| but with the user-defined macro-integrator replaced with second and fourth order, respectively, Runge--Kutta macro-integrators that take user specified macro-scale time-steps.

\subsection{Algorithms to simulate on patches of space}
\label{sec:justPat}
The spatial multiscale \ode{}s~\cref{eq:pat} are simulated only on a fraction of space by employing a patch scheme \cite[also known as the gap-tooth scheme,][]{Samaey08}.
This scheme applies to dynamic systems evolving in time with some `spatial' structure (`space' could be some other type of domain), and is useful when direct simulation over the spatial domain of interest is infeasible due to the overwhelming computational cost.
The toolbox assumes that space on the micro-scale is represented as a rectangular lattice network with grid points~\(\xv_i\)---one example being the spatial discretization~\cref{eq:pat} of a \pde.
The toolbox function requires the user to provide a function that computes a micro-scale time-step or time derivative for variables~\(\uv_i\) at each of \text{the lattice nodes.}

The patch scheme computes the detailed micro-scale at only a (small) subset of lattice nodes in space---the patches as illustrated by \cref{fig:patEx}. 
The scheme craftily couples the patches together to ensure macro-scale predictions are provably accurate \cite[]{Roberts06d, Roberts2011a, Bunder2013b, Cao2014a}.
For simplicity, this article discusses specifically the case of patches in 1D space, but the scheme and the toolbox extend to higher dimensional space \cite[]{Roberts2011a}, as in the 2D simulations of \cref{fig:patEx}. 
The essence of the toolbox patch algorithms are outlined, for a specific case in 1D~space, by \cref{alg:pat,alg:con,alg:int}.

\begin{algorithm}[btp]
\caption{\label{alg:pat}Patch scheme when a user has a function $\fun(t,\uv)$ that computes a time-step\slash time-derivative on a micro-scale lattice in space.
Suppose that direct simulation is prohibitively expensive with \fun\ by some time integrator, say \mac$(\fun, \{T_0, T\}, \uv_0)$.
A user then invokes the patch scheme with the following two steps.
}

%

\begin{algorithmic}[1]
 \State Invoke \Call{\cP}{$\fun, \{a,b\}, \ldots$} to configure the patch scheme on the spatial domain~\([a,b]\) of interest (\cref{alg:con}). 
 \State Then executing $\mac{(\protect\Call{\pS}{},\{T_0, T\}, \uv_0)}$ computes the solution on patches in space (\cref{alg:int}).
\end{algorithmic}
\end{algorithm}

\begin{algorithm}[btp]
\caption{\label{alg:con}
\cP\ initialises the basic patch structures for \cref{alg:pat}.
It creates \nPatch\ equi-spaced patches inside the 1D interval~\([a,b]\).  The patches are of size proportional to~\ratio\ with each having \nSubP\ micro-scale lattice points, and coupled by interpolation of order~\ordCC.
All defined variables are saved as global for subsequent use in simulation.}

\begin{algorithmic}[1]
\Function{\cP}{\fun, \{a,b\}, \nPatch, \ratio, \ordCC, \nSubP}
\State $H := (b-a)/N$; (distance between macro-scale mid-patches)
\State $X_j := a + (j-\frac12)H$ for $j:=1:\nPatch$; (macro-scale mid-patches)
\State $\io := ({\nSubP+1})/{2}$;  (index for the centre point of a patch)
\State $d := ({{\ratio}\times H})/({\io-1})$; (micro-scale lattice spacing)
\State \parbox[t]{0.9\linewidth}{\raggedright $x_{i,j} := X_j+ d(i-\io)$ for $j:=1:\nPatch$ and $i:=1:\nSubP$;  (micro-scale lattice points in the \(j\)th~patch) }
\State \parbox[t]{0.9\linewidth}{\raggedright Calculate coefficients of interpolation to determine the edge-values of a patch from mid-patch values, for interpolation from \(\ordCC/2\)~patches to either side of a patch. }  
\EndFunction
\end{algorithmic}

\end{algorithm}

\begin{algorithm}[btp]
\caption{\label{alg:int}
Evolve in time on patches via function \pS\ which interfaces between a user's \mac\ and \fun.  
Function \pS\ calculates the time-derivative\slash time-step in the interior of all patches via function \pEI\ which couples the patches with a macro-scale interpolation.
}
 

\begin{algorithmic}[1]
\Function{\pS}{$t$,$u$}
\State $u := \Call{\pEI}{u}$; 
\State $\dudt{}_{\text{interior}} := \fun(t,u,x)$; (evaluate time-step inside every patch)
\State $\dudt{}_{\text{edges},j}:=0$ for $j:=1:\nPatch$;  
\State $\dudt := \Call{reshape}{\dudt,\nSubP\times\nPatch,1}$; (make  column vector)
\State \Return $\dudt$
\EndFunction
\end{algorithmic}

\begin{algorithmic}[1]
\Function{\pEI}{$u$}
\State $u := \Call{reshape}{u,\nSubP, \nPatch}$; (reshape so each column is a patch)
\State \parbox[t]{0.9\linewidth}{\raggedright $u_{\text{edges},j} := {}$Lagrange interpolation of mid-patch values~\(u_{\scalebox{0.8}{\io},j}\) from \(\ordCC/2\)~patches to either side of each patch. }
\State \Return \(u\)
\EndFunction

\end{algorithmic}
\end{algorithm}

\cref{alg:pat} introduces the necessary code for using the toolbox to implement the patch scheme on a problem of interest. 
In the specific case of a macro-scale 1D domain~$[a,b]$, the toolbox function \cP\ (\cref{alg:con}) constructs \verb|nPatch| equi-spaced patches, each of width proportional to~\verb|ratio| and each containing \verb|nSubP| micro-scale lattice points.
The integer~\ordCC\ determines the form and order of the inter-patch coupling that ensures macro-scale accuracy: 
\begin{itemize}
\item even~\ordCC, the usual, and as described in \cref{alg:con}, invokes classic Lagrange interpolation, of order~\ordCC, from mid-patch values to the patch edge-values to generally give a macro-scale that is consistent to errors~\Ord{H^{\ordCC}} \cite[]{Roberts06d};
\begin{itemize}
\item the special case of \(\ordCC=0\), as used in 2D for the example of \cref{fig:patEx}, invokes a new spectral interpolation that recent numerical experiments indicate has consistency errors exponentially small in~\(H\);
\end{itemize}
\item  odd~\ordCC\ creates a scheme with a staggered grid of patches suitable for many wave systems \cite[]{Cao2013}, staggered in the sense that mid-patch values for odd/even patches use order~\ordCC\ interpolation to determine edge-patch values of even/odd patches, respectively, and that typically has macro-scales consistent to errors~\Ord{H^{\ordCC+1}} \cite[]{Cao2014a};
\begin{itemize}
\item and the special case of \(\ordCC=-1\) invokes a new staggered spectral interpolation that recent numerical experiments indicate has consistency errors exponentially small in~\(H\).
\end{itemize}
\end{itemize}
After constructing the patches, a user-specified integrator \mac\ (step~2 of \cref{alg:pat}) simulates the user-defined \verb|fun| over the time interval~$[T_0,T]$ with initial condition~$\vec{u}_0$. 
\cref{alg:int} overviews the toolbox function \pS\ which interfaces between the patch coupling and the user's function~\fun\ that computes the micro-scale time-derivative\slash time-step within all the patches.

The toolbox currently implements corresponding functionality for problems in 2D space, as \cref{fig:patEx} shows, via toolbox functions \verb|configPatches2|, \verb|patchSmooth2|, and \verb|patchEdgeInt2|.

This patch scheme is an example of so-called \emph{computational homogenization} \cite[e.g.,][]{Geers2010, Saeb2016, Geers2017} and is related to \emph{numerical homogenization} \cite[e.g.,][]{Craster2015, Owhadi2015, Peterseim2019, Maier2019}.
The three main distinguishing features of the patch scheme are: that a user need not perform any analysis of the micro-scale structures; that computations are done only on a (small) subset of the spatial domain; and for a wide class of systems the scheme is proved to be accurate to a user specified order \cite[]{Roberts06d, Roberts2011a, Cao2014a}.
However, in application to micro-scale heterogeneous media---the main interest of computational\slash numerical homogenization---more research needs to be done.
\cite{Bunder2013b} started exploring the patch scheme in heterogeneous media and established it generally has small macro-scale errors for diffusion in random media. 
Further, they found, analogous to that found in some numerical homogenization \cite[]{Peterseim2019}, that when the patch half-width is an integral number of periods of the micro-scale heterogeneity, then the macro-scale predictions are accurate to errors~\Ord{H^{\ordCC}}, as before.

A recent innovation in the toolbox (by setting parameter \verb|patches.EdgyInt=1|) is the capability to couple patches by interpolating next-to-edge values to the edge-patch values, but to the opposite edge.
This coupling is the subject of an article currently in preparation which will discuss how the coupling usefully preserves symmetry in many applications of interest, and is of controllable macro-scale accuracy for micro-scale heterogeneous media.

\section{Using toolbox functions}
\label{sec:tut}

Users need to download the toolbox via {GitHub}\footnote{\protect\url{https://github.com/uoa1184615/EquationFreeGit}}.
Place the folder of this toolbox in a path searched by your \script.
The toolbox provides both a user's and a developer's manual: start by looking at the User's Manual.

Many of the main toolbox functions, when invoked without any arguments, will simulate an example.
Executing the command \verb|PIG()| reproduces theProjective Integration example presented in \cref{sec:flagPI}.
Similarly, the nonlinear diffusion\slash lubrication-like example of \cref{sec:flagPat} is reproduced by executing \verb|configPatches2()|. 
The following \cref{sec:tutPI,sec:tutPat} explain the code for both of these introductory examples as templates to adapt for other problems.

\subsection{Invoking Projective Integration in General}
\label{sec:tutPI}

This subsection discusses some key factors when constructing a Projective Integration (PI) simulation.

\subsubsection{Burst must be long enough, and macro-time-steps short enough}
Suppose the slow dynamics of your system occurs at
rate\slash frequency of magnitude about~\(\alpha\); and the rate of \emph{decay} of your fast modes are
higher than the lower bound~\(\beta\) (e.g., if three fast
modes decay roughly like \(e^{-12t}, e^{-34t}, e^{-56t}\),
then \(\beta\approx 12\)). 
The PI must be able to stably project the (damped) fast modes, and so the duration~$\delta$ of the micro-scale burst must be sufficiently long. 
The fast modes decay like~$e^{-\beta\delta}$ over the micro-burst, and then grow like~$\beta\Delta$ on a projective step of length~$\Delta$. 
For stability, the product of these effects must be less than one; that is, $\beta\Delta e^{-\beta\delta} \lesssim 1$\,. 
Rearranging requires the burst length
\def\aD{\alpha\Delta}\def\bD{\beta\Delta}\def\dD{\delta/\Delta}
\begin{align}
\label{PIstab}\delta\gtrsim & \frac1{\beta}\log|\bD|
\end{align}
\cref{fig:bTlength} plots this lower bound as a function of~$\bD$ by writing $\dD\gtrsim\frac1{\bD}\log|\bD|$.

Once stability of the fast modes is assured by satisfying~\eqref{PIstab}, the accuracy of the macro-scale time-step must be considered.
The second and fourth order PI functions \texttt{PIRK2()} and \texttt{PIRK4()} have global errors proportional to~$\Delta^2$ and~$\Delta^4$ respectively, for a projective step of length~$\Delta$. 
More specifically, for a desired accuracy of \(\varepsilon\) and recalling that the slow dynamics occurs at rate\slash frequency of magnitude about~\(\alpha\): with \texttt{PIRK2} we suggest \(\alpha\Delta\lesssim(6\varepsilon)^{1/2}\);
whereas with \texttt{PIRK4} try \(\alpha\Delta\lesssim\varepsilon^{1/4}\).
The accuracy of \texttt{PIG()} is controlled by adjusting whatever the accuracy parameters are provided to the user-specified macro-integrator. 


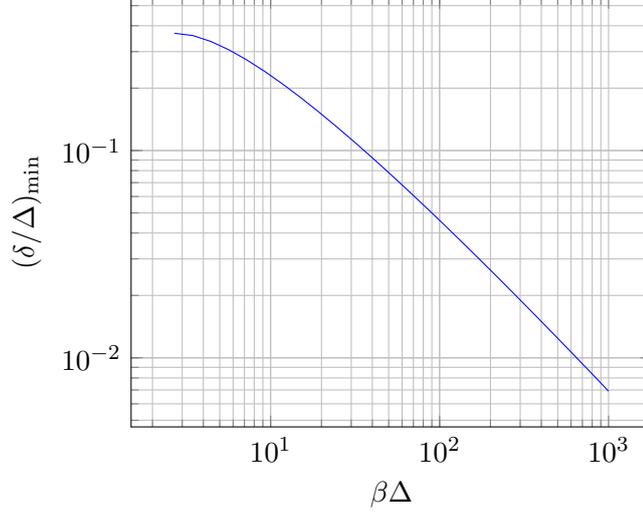
\begin{figure}
\centering
\caption{\label{fig:bTlength}Minimum value~\cref{PIstab} for~$\dD$, the ratio of burst length for the micro-simulator to the macro-time-step, in order to achieve stability in Projective Integration.
The horizontal axis~$\bD$ is the product of the minimum decay rate for the fast modes and the macro-time-step.
}
\tikzsetnextfilename{Figs/bTlength}
\begin{tikzpicture}
  \begin{loglogaxis}[xlabel={$\bD$}
  ,ylabel={$(\dD)_{\min}$}
  ,domain=2.7:1000 ,grid=both ]
  \addplot+[no marks]{ln(x)/x};
  \end{loglogaxis} 
\end{tikzpicture}
\end{figure}

\fancyvrbStartStop
\subsubsection{\texttt{PIG} tutorial}

We now discuss details of the simulation of the multiscale, slow-fast, \ode{}s~\cref{eq:sing} shown in \cref{fig:PIGsing}.
First we code the right-hand side function of the system (often people would phrase it in terms of the small parameter \(\epsilon=1/\beta\)),
\begin{matlab}
beta = 1e5;
dxdt=@(t,x) [ cos(x(1))*sin(x(2))*cos(t)
              beta*( cos(x(1))-x(2) ) ];
\end{matlab}
Second, we code micro-scale bursts, here using the standard
\verb|ode45()|.
We choose a burst length \((2/\beta) \log\beta\) as the fast rate of decay is approximately~\(\beta\).
Because we do not know the macro-scale time-step invoked by the adaptive function to be specified for \mac(), so we blithely assume \(\Delta\lesssim 1\) and then double the formula~\cref{PIstab} for safety. 
\begin{matlab}
bT = 2/beta*log(beta);
\end{matlab}
Then define the micro-scale burst from state~\verb|xb0| at time~\verb|tb0| to be an integration by the adaptive~\verb|ode45| of the coded \ode{}s~\cref{eq:sing}, over a burst-time~\verb|bT|.
\begin{matlab}
microBurst = @(tb0, xb0) feval('ode45',dxdt,[tb0 tb0+bT],xb0);
\end{matlab}
Third, code functions to convert between macro-scale state~$u_1$ and
micro-scale state~$(u_1,u_2)$.
\begin{matlab}
restrict = @(x) x(1);
lift = @(X,xApprox) [X; xApprox(2)];
\end{matlab}
The (optional) \texttt{restrict()} and \texttt{lift()} functions are detailed in the following \cref{sec:lift}. 
Fourth, invoke \verb|PIG| to use \Matlab's \verb|ode45| on the macro-scale slow evolution.
Integrate the micro-bursts over \(0\leq t\leq6\) from the initial condition \(\xv(0)=(1,0)\). 
\begin{matlab}
Tspan = [0 6]; 
x0 = [1;0];
macroInt='ode45';
[Ts,Xs,tms,xms] = PIG(macroInt,microBurst,Tspan,x0,restrict,lift);
\end{matlab}
We pause this example to discuss the available outputs from \verb|PIG()|.
Between zero and five outputs may be requested from \verb|PIG()|.
Most often you
would store the first two output results,
via say \verb|[Ts,Xs] = PIG(...)|.
\begin{itemize}
\item  \verb|T|, an \(L\)-vector of times at which
\mac()\ produced results.
\item \verb|X|, an \(L \times N\) array of the computed
solution: the \(i\)th~\emph{row} of~\verb|X|, \verb|X(i,:)|,
is to be the macro-state vector~\(\Xv(t_i)\) at time
\(t_i=\verb|T(i)|\).
\end{itemize}

However, micro-scale details of the underlying Projective
Integration computations may be helpful, and so \verb|PIG()|
provides some optional outputs of the micro-scale bursts, via
\verb|[Ts,Xs,tms,xms] = PIG(...)|

\begin{itemize}
\item \verb|tms|, optional, is an \(\ell\)-dimensional column
vector containing micro-scale times with bursts,
each burst separated by~\verb|NaN|; 

\item \verb|xms|, optional, is an \(\ell\times n\) array of
the corresponding micro-scale states. 
\end{itemize}
In some contexts it may be helpful to see directly how
Projective Integration approximates a reduced slow vector
field, via \verb|[T,X,tms,xms,svf] = PIG(...)| in which
\begin{itemize}
\item  \verb|svf| is a struct containing the
Projective Integration estimates of the slow vector field.
\begin{itemize}
\item \verb|svf.T| is a \(\hat L\)-dimensional  column
vector containing all times at which the micro-scale
simulation data is extrapolated to form an estimate of
\(d\xv/dt\) in \mac().
\item \verb|svf.dX| is a \(\hat L\times N\) array containing
the estimated slow vector field.
\end{itemize}
\end{itemize}
If \mac() is, for example, the forward Euler
method (or the Runge--Kutta method), then \(\hat L = L\) (or
\(\hat L = 4L \)). 

Returning to the example one remarkable feature of \verb|PIG()| is revealed: this PI simulation, which as mentioned in \cref{sec:flagPI} uses only~$0.6\%$ as many evaluations of~\cref{eq:sing} as direct simulation with \verb|ode45()|, is accomplished with a recursive call to \verb|ode45()|.
The standard \Matlab~integrator is used to both compute the micro-scale bursts, and also compute the projective steps.
All the usual machinery of adaptive time stepping and error control is used to regulate the micro-scale simulation, and also to regulate the macro-scale projective time-steps.

\subsubsection{Optional PI inputs enable the user to choose how to convert between macro- and micro-scale states}
\label{sec:lift}
As described in the penultimate paragraph of \cref{sec:justPI}, the user may require an intricate or application-specific process to convert between micro-scale variables~$u$ and macro-scale variables~$U$. 
Provide these bespoke functions to the toolbox PI functions with the optional inputs \verb|restrict()| and \verb|lift()|.
 
The user-provided function \verb}restrict(u)} should map a micro-scale state \verb}u}, at some fixed time, to a lower-dimensional macro-scale state \verb}U} at the same time. 
The reverse effect is accomplished by the function \verb}lift(U,uApprox)}, which takes two inputs---as lifting is often a non-unique process. 
The first input \verb|U| is the macro-scale state at the (present) time at which a micro-scale state is desired, and the second input \verb|uApprox| is the last micro-scale state output from the micro-simulator. 
This \verb|uApprox| is typically a micro-scale state from an earlier time, but nonetheless should be useful in initialising a consistent micro-scale state. 

A guiding principle in the restriction and lifting functions is that we cannot anticipate a user's every need; therefore, the functions are straightforward to edit.
In particular, their inputs may readily be expanded as needed. 

\subsubsection{Choose PI over stiff integrators in high dimensions}
\label{sec:choPI}
The 2D example of \cref{sec:flagPI} may be efficiently and simply simulated by standard stiff integrators, e.g.~\verb}ode15s()} in \Matlab. 
We here demonstrate the advantage of PI over such integrators as the model dimension increases. 
Consider linear systems of the form
\(\frac{d \uv}{dt} = {A}\uv + \bv \)\,,
where \((10+N)\times(10+N)\) matrix~${A}$ is randomly generated so that it has ten eigenvalues with real part within~$[-0.1,0.1]$, corresponding to ten slow variables, and $N$~eigenvalues with real part within $[-20\,000, -10\,000]$, corresponding to $N$~fast variables. 
The vector $\bv\in\RR^{10+N}$ is randomly generated with variance~one. 

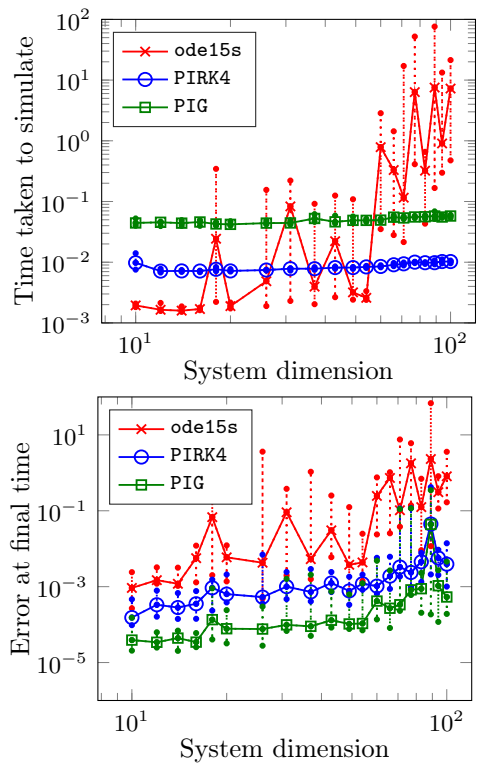
\begin{figure}
\centering 
\caption{\label{fig:piFor} 
Performance of Projective Integration compared to a standard stiff integrator, while scaling the dimension of the numerical model described in \cref{sec:choPI}. 
Solid lines show the median, and dotted lines show the 25th and 75th~percentiles. 
The stiff integrator is fast and reasonably accurate at low system dimension, but performs poorly at dimension~60 and above. }
    \tikzsetfigurename{Figs/PI4t}%
%
\begin{tikzpicture}

\begin{axis}[%
small, label shift={-1ex},
xmode=log,
xmin=0,
xmax=120,
xminorticks=true,
xlabel={System dimension},
ymode=log,
ymin=0.001,
ymax=100,
yminorticks=true,
ylabel={Time taken to simulate},
axis background/.style={fill=white},
legend style={at={(0.03,0.645)},font=\scriptsize, anchor=south west, legend cell align=left, align=left, draw=white!15!black}
]
\addplot [color=red, dotted, line width=0.7pt, mark size=0.8pt, mark=*, mark options={solid, red}, forget plot]
  table[row sep=crcr]{%
10	0.0019438745\\
10	0.0018222445\\
10	0.0020816215\\
10	0.0019438745\\
12	0.0016458115\\
12	0.001573342\\
12	0.0021351605\\
12	0.0016458115\\
14	0.0015923325\\
14	0.0015212905\\
14	0.001870956\\
14	0.0015923325\\
16	0.001693638\\
16	0.0016189595\\
16	0.0017939705\\
16	0.001693638\\
18	0.0243486545\\
18	0.0022209355\\
18	0.345045452\\
18	0.0243486545\\
20	0.0018912835\\
20	0.0017320815\\
20	0.0021648715\\
20	0.0018912835\\
26	0.004864205\\
26	0.001895348\\
26	0.1555478335\\
26	0.004864205\\
31	0.083114934\\
31	0.0023014835\\
31	0.2211064995\\
31	0.083114934\\
37	0.003963318\\
37	0.0020358665\\
37	0.0916518095\\
37	0.003963318\\
43	0.0221819915\\
43	0.0026536825\\
43	0.125226881\\
43	0.0221819915\\
49	0.0031880955\\
49	0.002413753\\
49	0.1089287125\\
49	0.0031880955\\
54	0.0025896285\\
54	0.0023964505\\
54	0.003339601\\
54	0.0025896285\\
60	0.785673057\\
60	0.0350706875\\
60	2.8488454905\\
60	0.785673057\\
66	0.330134546\\
66	0.0280223605\\
66	1.4343336065\\
66	0.330134546\\
71	0.115618427\\
71	0.021480852\\
71	17.091658476\\
71	0.115618427\\
77	6.337384126\\
77	0.411441199\\
77	52.119348237\\
77	6.337384126\\
83.0000000000001	0.3217392425\\
83.0000000000001	0.043193474\\
83.0000000000001	0.6578478425\\
83.0000000000001	0.3217392425\\
89	7.533309809\\
89	0.166863335\\
89	76.0829142565\\
89	7.533309809\\
94	0.9045407865\\
94	0.2978013435\\
94	13.2957480745\\
94	0.9045407865\\
100	7.252901275\\
100	0.475779869\\
100	21.370513688\\
100	7.252901275\\
};
\addplot [color=red, line width=0.7pt, mark size=2.5pt, mark=x, mark options={solid, red}]
  table[row sep=crcr]{%
10	0.0019438745\\
12	0.0016458115\\
14	0.0015923325\\
16	0.001693638\\
18	0.0243486545\\
20	0.0018912835\\
26	0.004864205\\
31	0.083114934\\
37	0.003963318\\
43	0.0221819915\\
49	0.0031880955\\
54	0.0025896285\\
60	0.785673057\\
66	0.330134546\\
71	0.115618427\\
77	6.337384126\\
83.0000000000001	0.3217392425\\
89	7.533309809\\
94	0.9045407865\\
100	7.252901275\\
};
\addlegendentry{\texttt{ode15s}}

\addplot [color=blue, dotted, line width=0.7pt, mark size=0.8pt, mark=*, mark options={solid, blue}, forget plot]
  table[row sep=crcr]{%
10	0.00980104\\
10	0.0075509785\\
10	0.0140542435\\
10	0.00980104\\
12	0.00714001\\
12	0.00676928\\
12	0.007811026\\
12	0.00714001\\
14	0.0071928205\\
14	0.006930785\\
14	0.0074429915\\
14	0.0071928205\\
16	0.007158974\\
16	0.0068587585\\
16	0.0072555385\\
16	0.007158974\\
18	0.007655408\\
18	0.0071113235\\
18	0.008306697\\
18	0.007655408\\
20	0.007186978\\
20	0.007020274\\
20	0.007697799\\
20	0.007186978\\
26	0.007435826\\
26	0.007280795\\
26	0.00751648\\
26	0.007435826\\
31	0.007820591\\
31	0.0072693185\\
31	0.00796193\\
31	0.007820591\\
37	0.007815824\\
37	0.0074900035\\
37	0.008286762\\
37	0.007815824\\
43	0.008154149\\
43	0.00780744\\
43	0.0084743275\\
43	0.008154149\\
49	0.008154739\\
49	0.007770126\\
49	0.00832729\\
49	0.008154739\\
54	0.008374362\\
54	0.0080466315\\
54	0.00869091749999999\\
54	0.008374362\\
60	0.00851138\\
60	0.008244004\\
60	0.0088997305\\
60	0.00851138\\
66	0.0090755665\\
66	0.00861224\\
66	0.009698001\\
66	0.0090755665\\
71	0.009409051\\
71	0.008850174\\
71	0.009881506\\
71	0.009409051\\
77	0.0100229805\\
77	0.0096744135\\
77	0.0101735035\\
77	0.0100229805\\
83.0000000000001	0.0098136435\\
83.0000000000001	0.009371465\\
83.0000000000001	0.0099635235\\
83.0000000000001	0.0098136435\\
89	0.0098240465\\
89	0.0092369615\\
89	0.010743711\\
89	0.0098240465\\
94	0.010255327\\
94	0.0095842375\\
94	0.0111304995\\
94	0.010255327\\
100	0.01026041\\
100	0.0096158495\\
100	0.010634558\\
100	0.01026041\\
};
\addplot [color=blue, line width=0.7pt, mark size=2.5pt, mark=o, mark options={solid, blue}]
  table[row sep=crcr]{%
10	0.00980104\\
12	0.00714001\\
14	0.0071928205\\
16	0.007158974\\
18	0.007655408\\
20	0.007186978\\
26	0.007435826\\
31	0.007820591\\
37	0.007815824\\
43	0.008154149\\
49	0.008154739\\
54	0.008374362\\
60	0.00851138\\
66	0.0090755665\\
71	0.009409051\\
77	0.0100229805\\
83.0000000000001	0.0098136435\\
89	0.0098240465\\
94	0.010255327\\
100	0.01026041\\
};
\addlegendentry{\texttt{PIRK4}}

\addplot [color=green!50!black, dotted, line width=0.7pt, mark size=0.8pt, mark=*, mark options={solid, green!50!black}, forget plot]
  table[row sep=crcr]{%
10	0.0443481865\\
10	0.039908693\\
10	0.05252737\\
10	0.0443481865\\
12	0.0454619325\\
12	0.0394124175\\
12	0.046652452\\
12	0.0454619325\\
14	0.044007387\\
14	0.0396093805\\
14	0.049363031\\
14	0.044007387\\
16	0.045965156\\
16	0.039712219\\
16	0.0494950885\\
16	0.045965156\\
18	0.0426576875\\
18	0.039832663\\
18	0.0457158725\\
18	0.0426576875\\
20	0.042043068\\
20	0.0398764225\\
20	0.0473572795\\
20	0.042043068\\
26	0.044085001\\
26	0.041851534\\
26	0.0488039265\\
26	0.044085001\\
31	0.0443365505\\
31	0.041573142\\
31	0.048651696\\
31	0.0443365505\\
37	0.052561637\\
37	0.0463675585\\
37	0.061823974\\
37	0.052561637\\
43	0.0462507855\\
43	0.043627764\\
43	0.0584494635\\
43	0.0462507855\\
49	0.048908938\\
49	0.0446127885\\
49	0.0565615095\\
49	0.048908938\\
54	0.0489280045\\
54	0.0459257285\\
54	0.0518836075\\
54	0.0489280045\\
60	0.0493690415\\
60	0.0464174355\\
60	0.052049764\\
60	0.0493690415\\
66	0.055244567\\
66	0.049671467\\
66	0.059209853\\
66	0.055244567\\
71	0.053678565\\
71	0.051425718\\
71	0.056956441\\
71	0.053678565\\
77	0.0556127055\\
77	0.0530874565\\
77	0.0580495985\\
77	0.0556127055\\
83.0000000000001	0.056107526\\
83.0000000000001	0.0524260715\\
83.0000000000001	0.059437147\\
83.0000000000001	0.056107526\\
89	0.0576486185\\
89	0.051644877\\
89	0.068752373\\
89	0.0576486185\\
94	0.0555169045\\
94	0.0537990775\\
94	0.059940297\\
94	0.0555169045\\
100	0.0576778585\\
100	0.053479626\\
100	0.0602052975\\
100	0.0576778585\\
};
\addplot [color=green!50!black, line width=0.7pt, mark size=1.8pt, mark=square, mark options={solid, green!50!black}]
  table[row sep=crcr]{%
10	0.0443481865\\
12	0.0454619325\\
14	0.044007387\\
16	0.045965156\\
18	0.0426576875\\
20	0.042043068\\
26	0.044085001\\
31	0.0443365505\\
37	0.052561637\\
43	0.0462507855\\
49	0.048908938\\
54	0.0489280045\\
60	0.0493690415\\
66	0.055244567\\
71	0.053678565\\
77	0.0556127055\\
83.0000000000001	0.056107526\\
89	0.0576486185\\
94	0.0555169045\\
100	0.0576778585\\
};
\addlegendentry{\texttt{PIG}}

\end{axis}
\end{tikzpicture}%
    \tikzsetfigurename{Figs/PI4e}%
%
\begin{tikzpicture}

\begin{axis}[%
small, label shift={-1ex},
xmode=log,
xmin=0,
xmax=120,
xminorticks=true,
xlabel={System dimension},
ymode=log,
ymin=1e-06,
ymax=100,
yminorticks=true,
ylabel={Error at final time},
legend style={at={(0.03,0.645)},font=\scriptsize,anchor=south west, legend cell align=left, align=left, draw=white!15!black}
]
\addplot [color=red, dotted, line width=0.7pt, mark size=0.8pt, mark=*, mark options={solid, red}, forget plot]
  table[row sep=crcr]{%
10	0.000925325761456185\\
10	0.000269256698267154\\
10	0.00240962395590021\\
10	0.000925325761456185\\
12	0.0014853087479113\\
12	0.0012000167117843\\
12	0.00322544008644119\\
12	0.0014853087479113\\
14	0.00119056734923021\\
14	0.00104875181604311\\
14	0.00316702908075387\\
14	0.00119056734923021\\
16	0.00563923004272277\\
16	0.00129839101625601\\
16	0.0120423604807888\\
16	0.00563923004272277\\
18	0.0677748010815422\\
18	0.00200799107899278\\
18	1.4459589884357\\
18	0.0677748010815422\\
20	0.00592479155571089\\
20	0.00137741242054833\\
20	0.0122465378267745\\
20	0.00592479155571089\\
26	0.00426935251571082\\
26	0.000484550064176811\\
26	3.6124476730972\\
26	0.00426935251571082\\
31	0.0911824594744185\\
31	0.0019731861099242\\
31	0.378921968080086\\
31	0.0911824594744185\\
37	0.00531027177653118\\
37	0.00157635576544794\\
37	1.06554017266145\\
37	0.00531027177653118\\
43	0.0305316837716183\\
43	0.0059925702519858\\
43	0.252041690194725\\
43	0.0305316837716183\\
49	0.00377211058093912\\
49	0.000707665093791869\\
49	0.124738974843011\\
49	0.00377211058093912\\
54	0.00435848380459048\\
54	0.00141643267310521\\
54	0.0248191350489678\\
54	0.00435848380459048\\
60	0.245946033607495\\
60	0.0238044959723955\\
60	0.756849432989389\\
60	0.245946033607495\\
66	0.732378487638789\\
66	0.0255143229791293\\
66	1.03627388819131\\
66	0.732378487638789\\
71	0.103494981721249\\
71	0.0380283435314088\\
71	7.55639809985078\\
71	0.103494981721249\\
77	1.78183059444223\\
77	0.139235593368984\\
77	6.03904078430665\\
77	1.78183059444223\\
83.0000000000001	0.12496254424896\\
83.0000000000001	0.00911172882767316\\
83.0000000000001	0.698605288867582\\
83.0000000000001	0.12496254424896\\
89	2.31192177750996\\
89	0.0116501051368253\\
89	68.4448178611952\\
89	2.31192177750996\\
94	0.308715599447179\\
94	0.114525142751932\\
94	0.814341856565296\\
94	0.308715599447179\\
100	0.810328743210349\\
100	0.167421796854314\\
100	3.5846846548374\\
100	0.810328743210349\\
};
\addplot [color=red, line width=0.7pt, mark size=2.5pt, mark=x, mark options={solid, red}]
  table[row sep=crcr]{%
10	0.000925325761456185\\
12	0.0014853087479113\\
14	0.00119056734923021\\
16	0.00563923004272277\\
18	0.0677748010815422\\
20	0.00592479155571089\\
26	0.00426935251571082\\
31	0.0911824594744185\\
37	0.00531027177653118\\
43	0.0305316837716183\\
49	0.00377211058093912\\
54	0.00435848380459048\\
60	0.245946033607495\\
66	0.732378487638789\\
71	0.103494981721249\\
77	1.78183059444223\\
83.0000000000001	0.12496254424896\\
89	2.31192177750996\\
94	0.308715599447179\\
100	0.810328743210349\\
};
\addlegendentry{\texttt{ode15s}}

\addplot [color=blue, dotted, line width=0.7pt, mark size=0.8pt, mark=*, mark options={solid, blue}, forget plot]
  table[row sep=crcr]{%
10	0.000153479565592456\\
10	9.6726575279992e-05\\
10	0.000464309521164511\\
10	0.000153479565592456\\
12	0.000333684634325165\\
12	0.00016194357737103\\
12	0.00078860036899624\\
12	0.000333684634325165\\
14	0.000281620495113463\\
14	0.000141351779117811\\
14	0.000678339973777285\\
14	0.000281620495113463\\
16	0.000354526215210887\\
16	0.000142328986754009\\
16	0.000765017024429468\\
16	0.000354526215210887\\
18	0.000931025914323591\\
18	0.0002341416509413\\
18	0.00153110638589726\\
18	0.000931025914323591\\
20	0.000637305602352156\\
20	0.000384252318497459\\
20	0.00209194374823238\\
20	0.000637305602352156\\
26	0.000530612721783212\\
26	0.000361616153135408\\
26	0.00699148534001698\\
26	0.000530612721783212\\
31	0.00100066186641595\\
31	0.000469571655384719\\
31	0.00243718926647054\\
31	0.00100066186641595\\
37	0.000727297169656117\\
37	0.000408577227931466\\
37	0.00291317534079633\\
37	0.000727297169656117\\
43.0000000000001	0.00124606970066933\\
43.0000000000001	0.000720653001299174\\
43.0000000000001	0.00242583650184191\\
43.0000000000001	0.00124606970066933\\
49	0.000772785585130036\\
49	0.000334277773665914\\
49	0.00182874612674521\\
49	0.000772785585130036\\
54.0000000000001	0.00119302306626327\\
54.0000000000001	0.000827597352448095\\
54.0000000000001	0.00207608760047172\\
54.0000000000001	0.00119302306626327\\
60	0.00103437620918857\\
60	0.000654601812087839\\
60	0.0057548585237309\\
60	0.00103437620918857\\
66.0000000000001	0.0018929574854371\\
66.0000000000001	0.000874250689395302\\
66.0000000000001	0.00613049126160497\\
66.0000000000001	0.0018929574854371\\
71	0.00329380464643746\\
71	0.00184860942264862\\
71	0.115129274226735\\
71	0.00329380464643746\\
77.0000000000001	0.00236549374282102\\
77.0000000000001	0.000969482192774739\\
77.0000000000001	0.117863847987931\\
77.0000000000001	0.00236549374282102\\
83	0.0045306148687294\\
83	0.00133119844203222\\
83	0.00780494255761078\\
83	0.0045306148687294\\
89.0000000000001	0.0447509988085789\\
89.0000000000001	0.0021012089406575\\
89.0000000000001	0.420130204272028\\
89.0000000000001	0.0447509988085789\\
93.9999999999999	0.00570419602321138\\
93.9999999999999	0.00202562596500593\\
93.9999999999999	0.00936993034910751\\
93.9999999999999	0.00570419602321138\\
100	0.00391787617877122\\
100	0.00100733546488652\\
100	0.0139133809001442\\
100	0.00391787617877122\\
};
\addplot [color=blue, line width=0.7pt, mark size=2.5pt, mark=o, mark options={solid, blue}]
  table[row sep=crcr]{%
10	0.000153479565592456\\
12	0.000333684634325166\\
14	0.000281620495113463\\
16	0.000354526215210887\\
18	0.000931025914323591\\
20	0.000637305602352156\\
26	0.000530612721783212\\
31	0.00100066186641595\\
37	0.000727297169656117\\
43	0.00124606970066933\\
49	0.000772785585130036\\
54	0.00119302306626327\\
60	0.00103437620918857\\
66	0.0018929574854371\\
71	0.00329380464643746\\
77	0.00236549374282102\\
83.0000000000001	0.0045306148687294\\
89	0.0447509988085789\\
94	0.00570419602321138\\
100	0.00391787617877123\\
};
\addlegendentry{\texttt{PIRK4}}

\addplot [color=green!50!black, dotted, line width=0.7pt, mark size=0.8pt, mark=*, mark options={solid, green!50!black}, forget plot]
  table[row sep=crcr]{%
10	3.90831001681369e-05\\
10	2.06175922342733e-05\\
10	0.000152693452599978\\
10	3.90831001681369e-05\\
12	3.42514604656476e-05\\
12	2.48086592843389e-05\\
12	6.27415104081127e-05\\
12	3.42514604656476e-05\\
14	4.40708131786107e-05\\
14	2.03437234850179e-05\\
14	6.92380330484766e-05\\
14	4.40708131786107e-05\\
16	3.45302523885196e-05\\
16	2.54582761541003e-05\\
16	5.98703183243855e-05\\
16	3.45302523885196e-05\\
18	0.000135068051272942\\
18	4.05534951006008e-05\\
18	0.000905233844990876\\
18	0.000135068051272942\\
20	7.78656274080677e-05\\
20	3.20268692272127e-05\\
20	0.000240430133370868\\
20	7.78656274080677e-05\\
26	7.63958436151795e-05\\
26	2.78113929216013e-05\\
26	0.000298288793935635\\
26	7.63958436151795e-05\\
31	9.8325094161416e-05\\
31	6.81917681415764e-05\\
31	0.00160443285611512\\
31	9.8325094161416e-05\\
37	9.00610730926877e-05\\
37	5.04863714923781e-05\\
37	0.00205799577866294\\
37	9.00610730926877e-05\\
43.0000000000001	0.000130305153236655\\
43.0000000000001	8.35767473095796e-05\\
43.0000000000001	0.000748762578458527\\
43.0000000000001	0.000130305153236655\\
49	0.000104072497112474\\
49	7.67658444504513e-05\\
49	0.000624351996100516\\
49	0.000104072497112474\\
54.0000000000001	0.000107002652161126\\
54.0000000000001	7.07740992466916e-05\\
54.0000000000001	0.00122040260640267\\
54.0000000000001	0.000107002652161126\\
60	0.000417245285025704\\
60	0.000136879708652034\\
60	0.0048336779879597\\
60	0.000417245285025704\\
66.0000000000001	0.000269941093856936\\
66.0000000000001	8.14911245085095e-05\\
66.0000000000001	0.00267333962997086\\
66.0000000000001	0.000269941093856936\\
71	0.000334013859892547\\
71	0.000230155212671826\\
71	0.111298765348944\\
71	0.000334013859892547\\
77.0000000000001	0.000823664548835088\\
77.0000000000001	0.000603799916896137\\
77.0000000000001	0.115782590047759\\
77.0000000000001	0.000823664548835088\\
83	0.000883319619288055\\
83	0.000205150632960139\\
83	0.00229994887996827\\
83	0.000883319619288055\\
89.0000000000001	0.0436708630852678\\
89.0000000000001	0.000186504543684881\\
89.0000000000001	0.348279521521476\\
89.0000000000001	0.0436708630852678\\
93.9999999999999	0.00105781230536829\\
93.9999999999999	0.00011797445315255\\
93.9999999999999	0.00271199655473271\\
93.9999999999999	0.00105781230536829\\
100	0.000534135190851425\\
100	0.000191540395798047\\
100	0.00457263293489575\\
100	0.000534135190851425\\
};
\addplot [color=green!50!black, line width=0.7pt, mark size=1.8pt, mark=square, mark options={solid, green!50!black}]
  table[row sep=crcr]{%
10	3.90831001681369e-05\\
12	3.42514604656476e-05\\
14	4.40708131786107e-05\\
16	3.45302523885196e-05\\
18	0.000135068051272942\\
20	7.78656274080677e-05\\
26	7.63958436151795e-05\\
31	9.8325094161416e-05\\
37	9.00610730926877e-05\\
43.0000000000001	0.000130305153236655\\
49	0.000104072497112474\\
54.0000000000001	0.000107002652161126\\
60	0.000417245285025704\\
66.0000000000001	0.000269941093856936\\
71	0.000334013859892547\\
77.0000000000001	0.000823664548835088\\
83	0.000883319619288055\\
89.0000000000001	0.0436708630852678\\
93.9999999999999	0.00105781230536829\\
100	0.000534135190851425\\
};
\addlegendentry{\texttt{PIG}}

\end{axis}
\end{tikzpicture}%
     
\end{figure}%
We generate such a system at \(20\)~values of~$N$ between~\(0\) and~\(90\), and at each one simulate to final time~\(10\) from randomly chosen initial conditions with each of \verb}PIRK4()}, \verb}PIG()} and \verb}ode15s()}. 
After each simulation we record the time elapsed over each simulation as well as the relative error between the simulation estimates and the exact solution. 
This procedure is repeated a further eleven times (the toolbox script \verb|pirk4mance.m| details code for this experiment): \cref{fig:piFor} displays the recorded statistics. 
It appears that for system dimension larger than about~\(60\), the cost by \verb|ode15s| in setting up and managing the Jacobian is too high.
For these larger systems, the Projective Integration functions appear the better choice.

The relative performance of \verb}ode15s} may be improved by providing it with more information. 
By providing both the Jacobian~${A}$ explicitly to the integrator, and specifying the initial conditions on the attracting slow manifold, \verb|ode15s| becomes competitive with the Projective Integration functions (up to system dimension~$190$). 
But knowing and coding both the system Jacobian and slow manifold is usually too hard in practice.

\subsection{Patch dynamics tutorial}
\label{sec:tutPat}

The code here shows one way to get started with the patch scheme. 
A user's script may have the following three steps (arrows indicate function recursion).
\begin{enumerate}\def\itemsep{-1.0ex}
\item configPatches2 
\item ode integrator \into patchSmooth2 \into user's micro-scale code
\item process results
\end{enumerate}
We reproduce the simulation (\cref{fig:patEx}) of the lattice spatial discretization~\cref{eq:pat} of a nonlinear diffusion \pde.
As a micro-scale discretization of the \pde~we use the following function to compute the time derivatives: an array variable \verb|u(i,j,:,:)| refers to the \((i,j)\)th~point in each and every patch as the third and forth indices index the 2D array of patches.
\begin{matlab}
function ut = nonDiffPDE(t,u,x,y)
  dx = diff(x(1:2));  dy = diff(y(1:2));  
  i = 2:size(u,1)-1;  j = 2:size(u,2)-1;  
  ut = nan(size(u));  
  ut(i,j,:,:) = diff(u(:,j,:,:).^3,2,1)/dx^2 ...
               +diff(u(i,:,:,:).^3,2,2)/dy^2;
end
\end{matlab}

To use this micro-scale code on patches we need to establish global data, in  struct~\verb|patches|, to characterise the patches.
Here we aim to simulate the nonlinear diffusive~\cref{eq:pat}
on a macro-scale \(6\times4\)-periodic domain with a 2D array of \(9\times7\) patches.
Choose spectral interpolation~(\(\ordCC = 0\)) to couple the patches, and set each
patch of half-size ratio~\(0.25\) (relatively large for
visualization), and with \(5\times5\) micro-scale lattice points within each
patch. 
\cite{Roberts2011a} established that such a patch scheme is
consistent with the discretisation~\cref{eq:pat}, as the patch spacing~\(H\) decreases, and hence is consistent to the original~\pde.
\begin{matlab}
global patches
nSubP = 5;
configPatches2(@nonDiffPDE,[-3 3 -2 2],nan,[9 7],0,0.25,nSubP);
\end{matlab}
The third argument to \verb|configPatches2()| is intended for boundary conditions on the macro-scale simulation, not yet implemented in the toolbox.
The inputs are otherwise 2D analogues of the inputs to \cP\ in \cref{alg:pat}.

For an initial condition of the simulation, here set a perturbed-Gaussian.
The \verb|reshape| functions give the \(x\) and~\(y\) lattice coordinates as two 4D arrays of size \(5\times1\times9\times1\) and \(1\times5\times1\times7\) respectively.
Then auto-replication fills~\verb|u0| with values from the Gaussian \(u=e^{-x^2-y^2}\).
\begin{matlab}
x = reshape(patches.x,nSubP,1,[],1); 
y = reshape(patches.y,1,nSubP,1,[]);
u0 = exp(-x.^2-y.^2);
u0 = u0.*(0.9+0.1*rand(size(u0)));
\end{matlab}
Initiate a plot of the simulation using only the micro-scale
values interior to the patches: set \(x\)~and \(y\)-edges to
\verb|nan| to leave, in plots, gaps between patches. 
\begin{matlab}
figure(1), clf
x = patches.x; y = patches.y;
x([1 end],:) = nan; y([1 end],:) = nan;
\end{matlab}
Start by showing the initial conditions---the top-left panel of \cref{fig:patEx}.
\begin{matlab}
u = reshape(permute(u0,[1 3 2 4]), [numel(x) numel(y)]);
hsurf = surf(x(:),y(:),u'); drawnow
\end{matlab}
Integrate in time using a standard function, or alternatively via projective integration.
\begin{matlab}
[ts,us] = ode15s( @patchSmooth2, [0 4], u0(:));
\end{matlab}
Animate the computed simulation to finish with the bottom-right picture in \cref{fig:patEx}.
Use \verb|patchEdgeInt2| to interpolate patch-edge values (even if not drawn) as it reconstitutes the row orientated output from \verb|ode15s| into arrays that reflect the 2D patch structure.
\begin{matlab}
for i = 1:length(ts)
  u = patchEdgeInt2(us(i,:));
  u = reshape(permute(u,[1 3 2 4]), [numel(x) numel(y)]);
  set(hsurf,'ZData', u');
  legend(['time = ' num2str(ts(i),2)])
  pause(0.1)
end
\end{matlab}
%

\section{Discussion}
\label{sec:Disc}

This project is collectively developing a \script\ toolbox of equation-free algorithms \cite[]{Roberts2019b}.
The algorithms currently implement a useful functionality, but much more is desirable so the plan is to subsequently develop more capability.
\script\ appears a good choice for a first version since it is widespread, efficient, supports various parallel modes, and development costs are reasonably low.
Further, it is built on \textsc{blas} and \textsc{lapack} so the cache and superscalar \cpu{} are potentially \text{well utilised.}

Projective Integration and Patch Scheme functions are designed to generally enable users to invoke `equation-free' algorithms in a wide variety of applications.
In addition to simulations analogous to those in previous sections,  the toolbox may empower users in other scenarios such as the following list.
In the User Manual \cite[]{Roberts2019b} each toolbox function and application is presented in its own section, so here we refer to sections of the User Manual by using the name of the appropriate function. 
\begin{itemize}\sloppy
\item To projectively integrate in time a multiscale, slow-fast, system of \ode{}s with a \emph{simple} PI macro-integrator, you could use \verb|PIRK2()|, or \verb|PIRK4()| for higher-order accuracy. 
Perhaps adapt the Michaelis--Menten example presented in the User Manual at the beginning of~\verb|PIRK2.m|.

\item One may use short forward bursts of micro-scale simulation in order to stably predict the slow dynamics \emph{backward in time} using PI \cite[]{Gear03b}, as in \verb|egPIMM.m|.

\item The lifting and restriction functions may be utilised in more complicated applications like those others have previously addressed \cite[e.g.,][]{Frederix2007, Roose2009, Bold2012, Sieber2018}.
The lifting function has two inputs: the macro-scale state (after a macro-scale time step); and the micro-scale state at the end of the last micro-simulation.
Full details for constructing these functions are in \verb|PIG.m|. 

\item For the patch scheme in 1D adapt the code at the beginning of \verb|configPatches1.m| for Burgers' \pde, or the staggered patches of 1D water wave equations in \verb|waterWaveExample.m|.

\item For the patch scheme in 2D adapt the code at the beginning of \verb|configPatches2.m| for nonlinear diffusion, or the regular patches of the 2D wave equation of \verb|wave2D.m|.

\item The previous two examples are for systems that have \emph{smooth} spatial structures on the micro-scale: when the micro-scale is `rough' with a known heterogeneous period (so far only in 1D), then adapt 
the example of \verb|HomogenisationExample.m|.

\item Employ an ensemble of patch dynamics simulations, averaging appropriately, by adapting the example of \verb|ensembleAverageExample.m|.

\item Combine the projective integration and patch functions to simulate a system with both time and spatial scale separation.
In this case use a PI function as the integrator for a patch scheme, as done in the second portion of \verb|ensembleAverageExample.m|. 
\end{itemize}

We encourage adaptation and further development of the toolbox algorithms, and are keen to include new collaborators in future versions of the toolbox. 
In particular, as well as developing multi-D functions further, corresponding 1D coded functionality, we need to code and prove capability for general macro-scale boundaries.
We also plan to develop projective integration for oscillatory\slash stochastic systems, perhaps via Dynamic Mode Decomposition \cite[e.g.,][]{Kutz2016, Kutz2018}.

\paragraph{Acknowledgements} 
This project is supported by the Australian Research Council via grants DP150102385 and DP180100050.

\bibliography{bibexport}

\end{document}